\documentstyle[12pt]{article}

\font\twlgot =eufm10 scaled \magstep1
\font\egtgot =eufm8
\font\sevgot =eufm7

\font\twlmsb =msbm10 scaled \magstep1
\font\egtmsb =msbm8
\font\sevmsb =msbm7

\newfam\gotfam
\def\pgot{\fam\gotfam\twlgot}
\textfont\gotfam\twlgot
\scriptfont\gotfam\egtgot
\scriptscriptfont\gotfam\sevgot
\def\got{\protect\pgot}

\newfam\msbfam
\textfont\msbfam\twlmsb
\scriptfont\msbfam\egtmsb
\scriptscriptfont\msbfam\sevmsb

\def\pBbb{\relax\ifmmode\expandafter\Bb\else\typeout{You cann't use
Bbb in text mode}\fi}
\def\Bb #1{{\fam\msbfam\relax#1}}

\def\thebibliography#1{\bigskip\section*{\large
\bf References\\}\list
  {[\arabic{enumi}]}{\settowidth\labelwidth{#1}\leftmargin\labelwidth
    \advance\leftmargin\labelsep
    \usecounter{enumi}}
    \def\newblock{\hskip .11em plus .33em minus .07em}
    \sloppy\clubpenalty4000\widowpenalty4000
    \sfcode`\.=1000\relax}

\def\op#1{\mathop{{\it\fam0} #1}\limits}

\newcommand{\di}{{\rm dim\,}}

\newcommand{\beq}{\begin{equation}}
\newcommand{\eeq}{\end{equation}}
\newcommand{\ben}{\begin{eqnarray}}
\newcommand{\een}{\end{eqnarray}}
\newcommand{\be}{\begin{eqnarray*}}
\newcommand{\ee}{\end{eqnarray*}}
\newcommand{\bea}{\begin{eqalph}}
\newcommand{\eea}{\end{eqalph}}

\newcommand{\cG}{{\got g}}

\newcommand{\gS}{{\got S}}

\newcommand{\gF}{{\got F}}

\newcommand{\cL}{{\cal L}}

\newcommand{\cF}{{\cal F}}

\newcommand{\bL}{{\bf L}}

\newcommand{\al}{\alpha}
\newcommand{\bt}{\beta}

\newcommand{\la}{\lambda}
\newcommand{\La}{\Lambda}
\newcommand{\f}{\phi}

\newcommand{\m}{\mu}

\newcommand{\g}{\gamma}

\newcommand{\ve}{\varepsilon}

\newcommand{\vt}{\vartheta}

\newcommand{\si}{\sigma}

\newcommand{\w}{\wedge}

\newcommand{\dr}{\partial}

\newcommand{\ot}{\otimes}

\newenvironment{eqalph}{\stepcounter{equation}
\setcounter{equationa}{\value{equation}}
\setcounter{equation}{0}

\begin{eqnarray}}{\end{eqnarray}\setcounter{equation}{\value{equationa}}}

\newcounter{example}
\newcounter{remark}
\newcounter{theorem}
\newcounter{proposition}
\newcounter{lemma}
\newcounter{corollary}
\newcounter{definition}

\def\theremark{\arabic{remark}}

\def\thedefinition{\arabic{definition}}

\newcommand{\bs}{{\bf s}}

\newcommand{\mar}[1]{}

\begin{document}
\hbox{}


\begin{center}

{\large \bf The BRST extension of gauge non-invariant Lagrangians}  
\bigskip 

{\sc D.Bashkirov\footnote{{\it E-mail address}: bashkird@rol.ru}
 and G.Sardanashvily\footnote{{\it E-mail address}:
sard@grav.phys.msu.su; \,\, {\it Web}:
http://webcenter.ru/$\sim$sardan/}} 

\medskip

\begin{small}

Department of Theoretical Physics, Moscow State University, 117234
Moscow, Russia
\medskip
\end{small}

\end{center}

\begin{small}

\noindent 
{\bf Abstract.}
We show that, in gauge theory of principal connections, any gauge non-invariant
Lagrangian can be completed to the BRST-invariant one. The BRST
extension of the global Chern--Simons Lagrangian is present.

\end{small}

\bigskip
\bigskip

In perturbative quantum gauge theory, the BRST symmetry has been found
as a symmetry of the gauge fixed Lagrangian \cite{fadd}. The
ghost-free summand $L_1$ of this Lagrangian is gauge non-invariant, but it is
completed to the BRST-invariant Lagrangian
$L=L_1+L_2$ by means of the term $L_2$ depending on ghosts and anti-ghosts.
We aim to show that any gauge non-invariant Lagrangian can be extend to
the BRST-invariant one, though the anti-ghost sector of this BRST
symmetry differs from that in \cite{fadd}.

In a general setting, let us consider a Lagrangian BRST model with a
nilpotent odd BRST operator $\bs$ of ghost number 1. Let $L$ be a
Lagrangian of zero ghost number which need not be BRST-invariant, 
i.e., $\bs L\neq 0$.
Let us complete the physical basis of this BRST model with an odd anti-ghost 
field $\si$ of ghost number $-1$. Then we introduce the modified
BRST operator 
\mar{br1}\beq
\bs'=\frac{\dr}{\dr\si} +\bs \label{br1}
\eeq
which is also a nilpotent odd operator of 
ghost number 1. Let us consider the Lagrangian
\mar{br2}\beq
L'=\bs'(\si L)=L-\si\bs L. \label{br2}
\eeq 
Since $\bs$ is nilpotent, this Lagrangian is $\bs'$-invariant, i.e.,
$\bs'L'=0$. Moreover, it is readily observed that any $\bs'$-invariant
Lagrangian takes the form (\ref{br2}). It follows that the cohomology
of the BRST operator (\ref{br1}) is trivial.

Turn now to the gauge theory of principal connections on a principal
bundle $P\to X$ with a structure Lie group $G$. Let $VP$ and $J^1P$
denote the vertical tangent bundle and the first order jet manifold of 
$P\to X$, respectively. Principal connections on $P\to X$ are
represented by sections of  the affine bundle
\mar{br3}\beq
C=J^1P/G\to X, \label{br3}
\eeq
modelled over the vector bundle $T^*X\ot V_GP$ \cite{book00}. Here,
$V_GP=VP/G$ is the fibre bundle in Lie algebras $\cG$ of the group $G$.
Given the basis $\{\ve_r\}$ for $\cG$, we obtain the local fibre bases
$\{e_r\}$ for $V_GP$. There is one-to-one correspondence between the
sections $\xi=\xi^r e_r$ of $V_GP\to X$ and the vector fields on $P$ which
are infinitesimal generators of one-parameter groups of vertical
automorphisms (i.e., gauge
transformations) of $P$. The connection bundle $C$ (\ref{br3}) is
coordinated by $(x^\m,a^r_\m)$ such that, written
relative to these coordinates, sections 
$A=A^r_\m dx^\m\ot e_r$ of $C\to X$ are the familiar
local connection one-forms, regarded as gauge potentials. 
The
configuration space of these gauge potentials is the infinite order jet
manifold $J^\infty C$ coordinated by $(x^\m, a^r_\m, a^r_{\La\m})$,
$0<|\La|$, where $\La=(\la_1\cdots\la_k)$, $|\La|=k$, denotes a
symmetric multi-index. A $k$-order Lagrangian of gauge potentials
is given by a horizontal density
\mar{br4}\beq
L=\cL(x^\m,a^r_\m, a^r_{\La\m})d^nx, \qquad 0<|\La|\leq k, \qquad n=\di
X, \label{br4}
\eeq 
of jet order $k$ on $J^\infty C$. 

Any section $\xi=\xi^r e_r$ of the
Lie algebra bundle $V_GP\to X$ yields the vector field 
\mar{br6}\beq
u_\xi=u^r_\m\frac{\dr}{\dr a^r_\m}=(\dr_\m\xi^r +c^r_{pq}a^p_\m\xi^q)
\frac{\dr}{\dr a^r_\m} \label{br6}
\eeq
on $C$ where $c^r_{pq}$ are the structure constants of the Lie algebra
$\cG$. This vector field is the
infinitesimal generator of
a one-parameter group of gauge transformations of $C$.
Its prolongation onto the configuration space $J^\infty C$ reads
\mar{br5,7}\ben
&& J^\infty u_\xi=u_\xi +\op\sum_{0<|\La|}d_\La
u^r_\m\frac{\dr}{\dr a^r_{\La\m}}, \label{br5}\\
&& d_\La=d_{\la_1}\cdots d_{\la_k}, \qquad
d_\la=\dr_\la +a^r_{\la\m}\frac{\dr}{\dr a^r_\m} +
a^r_{\la\la_1\m}\frac{\dr}{\dr a^r_{\la_1\m}} +
a^r_{\la\la_1\la_2\m}\frac{\dr}{\dr a^r_{\la_1\la_2\m}}+\cdots. \label{br7}
\een
A Lagrangian $L$ (\ref{br4}) is called gauge-invariant iff its Lie derivative
\mar{br8}\beq
\bL_{J^\infty u_\xi}L= J^\infty u_\xi\rfloor d\cL d^nx=
(u_\xi +\op\sum_{0<|\La|}d_\La
u^r_\m\frac{\dr}{\dr a^r_{\La\m}})\cL d^nx \label{br8}
\eeq
along the vector field (\ref{br5}) vanishes for all infinitesimal gauge
transformations $\xi$.

Let us extend gauge theory on a principal bundle $P$ to a BRST model,
similar to that in \cite{barn,bran01}. Its physical basis consists of
polynomials in fibre coordinates $a^r_{\La\m}$, $|0\leq\La|$,
on $J^\infty C$ and the
odd elements $C^r_\La$, $|0\leq\La|$, of ghost number 1 which make up the local basis 
for the graded
manifold determined by the infinite order jet bundle $J^\infty V_GP$
\cite{mpl,epr2}. The BRST operator in this model is defined as the Lie
derivative 
\mar{br10}\beq
\bs=\bL_\vt \label{br10}
\eeq
along the graded
vector field
\mar{br9}\ben
&& \vt=v^r_\m\frac{\dr}{\dr a^r_\m} + \op\sum_{0<|\La|}d_\La
v^r_\m\frac{\dr}{\dr a^r_{\La\m}} + v^r\frac{\dr}{\dr C^r}+
\op\sum_{0<|\La|}d_\La
v^r\frac{\dr}{\dr C^r_\La}, \label{br9}\\
&& v^r_\m=C^r_\m +c^r_{pq}a^p_\m C^q, \qquad v^r=-\frac12 c^r_{pq}
C^pC^q, \nonumber
\een
where $d_\La$ is the generalization of the total derivative (\ref{br7})
such that
\mar{br15}\beq
d_\la=\dr_\la +[a^r_{\la\m}\frac{\dr}{\dr a^r_\m} +
a^r_{\la\la_1\m}\frac{\dr}{\dr a^r_{\la_1\m}} +\cdots]
+ [C^r_\la\frac{\dr}{\dr C^r} +
C^r_{\la\la_1}\frac{\dr}{\dr C^r_{\la_1}}+\cdots]. \label{br15}
\eeq
A direct computation shows that the operator $\bs$ (\ref{br10}) acting on horizontal
(local in the terminology of \cite{barn}) forms
\be
\f=\frac1{k!}\f_{\al_1\ldots\al_k}dx^{\al_1} \w\cdots\w dx^{\al_k}
\ee
is nilpotent, i.e.,
\be
\bL_\vt\bL_\vt\f=\vt\rfloor d(\vt\rfloor d\f)=[
\op\sum_{0\leq|\La|}\vt(d_\La
v^r_\m)\frac{\dr}{\dr a^r_{\La\m}} + 
\op\sum_{0\leq|\La|}\vt(d_\La
v^r)\frac{\dr}{\dr C^r_\La}]\f=0.
\ee

Let $L$ (\ref{br4}) be a (higher-order) Lagrangian of gauge theory. The
BRST operator $\bs$ (\ref{br10}) acts on $L$ as follows:
\be
\bs L= (v^r_\m\frac{\dr}{\dr a^r_\m} + \op\sum_{0<|\La|}d_\La
v^r_\m\frac{\dr}{\dr a^r_{\La\m}})\cL d^nx, \qquad 
v^r_\m=C^r_\m +c^r_{pq}a^p_\m C^q,
\ee
Comparing this expression with the expressions (\ref{br6}) and
(\ref{br8}) shows that a Lagrangian $L$ is gauge-invariant iff it is
BRST-invariant. If $L$ need not be gauge-invariant, one can follow the above
mentioned procedure of its BRST extension. Let us introduce the
anti-ghost field $\si$ and the modified BRST operator $\bs'$ (\ref{br1}).
Then the Lagrangian
\mar{br11}\beq
L'=L-\si\bs L=L-\si(v^r_\m\frac{\dr}{\dr a^r_\m} + \op\sum_{0<|\La|}d_\La
v^r_\m\frac{\dr}{\dr a^r_{\La\m}})\cL d^nx \label{br11}
\eeq
is $\bs'$-invariant. If $L$ is gauge-invariant, then $L'=L$. In
particular, let $L$ be 
a first order Lagrangian. Then its BRST extension (\ref{br11}) reads
\mar{br12}\beq
L'=L-\si[(C^r_\m +c^r_{pq}a^p_\m C^q)\frac{\dr}{\dr a^r_\m} +
(C^r_{\la\m} +c^r_{pq}a^p_{\la\m} C^q +c^r_{pq}a^p_\m C^q_\la) 
\frac{\dr}{\dr a^r_{\la\m}}]\cL d^nx. \label{br12}
\eeq

For example, let us obtain the BRST-invariant extension of the global
Chern--Simons Lagrangian. Let the structure group $G$ of a principal bundle
$P$ be semi-simple, and let $a^G$ be
the Killing form on $\cG$.
The connection bundle $C\to X$ (\ref{br3}) admits the canonical 
$V_GP$-valued 2-form
\be
\gF=(da^r_\m\w dx^\m +\frac12 c^r_{pq}a^p_\la a^q_\m dx^\la\w dx^\m)\ot
e_r.
\ee
Given a section $A$ of $C\to X$, the pull-back
\be
F_A=A^*\gF=\frac12 F(A)^r_{\la\m}dx^\la\w dx^\m\ot e_r,
\qquad
F(A)^r_{\la\m}=\dr_\la A^r_\m-\dr_\m A^r_\la +c^r_{pq}A^p_\la
A^q_\m,
\ee
of $\gF$ onto $X$ is the strength form of a gauge potential $A$.
Let
\be
P(\gF)=\frac{h}{2}a^G_{mn}\gF^m\w \gF^n 
\ee
be the second Chern characteristic form up to a constant multiple. 
Given a section $B$ of
$C\to X$, the corresponding global Chern--Simons three-form $\gS_3(B)$ on $C$
is defined by the transgression formula
\be
P(\gF)-P(F_B)=d\gS_3(B)
\ee
\cite{mpl2}. Let us consider the gauge model on a three-dimensional base
manifold $X$ with the global Chern--Simons Lagrangian 
\be
&&L_{\rm CS}=h_0(\gS_3(B))=
[\frac12ha^G_{mn} \ve^{\al\bt\g}a^m_\al(\cF^n_{\bt\g} -\frac13
c^n_{pq}a^p_\bt a^q_\g)  \\
&& \qquad -\frac12ha^G_{mn}
\ve^{\al\bt\g}B^m_\al(F(B)^n_{\bt\g} -\frac13 c^n_{pq}B^p_\bt
B^q_\g) -d_\al(ha^G_{mn} \ve^{\al\bt\g}a^m_\bt B^n_\g)]d^3x,
\ee
where $h_0(da^r_\m)=a^r_{\la\m}dx^\la$ and 
\be
\cF=h_0\gF=\frac12 \cF^r_{\la\m}dx^\la\w
dx^\m\ot e_r, \qquad
\cF^r_{\la\m}=a^r_{\la\m}-a^r_{\m\la} +c^r_{pq}a^p_\la a^q_\m.
\ee
This Lagrangian is globally defined, but it is not gauge-invariant
because of a background gauge potential $B$. Its BRST-invariant extension
(\ref{br12}) reads
\be
L'_{\rm CS}=L_{\rm CS} +ha^G_{mn}\si d_\al(\ve^{\al\bt\g}(C^m_\bt a^n_\g +
(C^m_\bt +c^m_{pq}a^p_\bt C^q) B^n_\g))d^3x,
\ee
where $d_\al$ is the total derivative (\ref{br15}).

\end{document}